\documentclass[prb,aps,twocolumn,amsmath,amssymb,floatfix,
superscriptaddress]{revtex4}

\usepackage[dvips]{graphics}
\usepackage{color}
\definecolor{dred}{rgb}{0,0,0.6}

\date{\today}

\begin{document}

\title{Engineering magnetoresistance: A new perspective}

\author{Moumita Patra}

\affiliation{Physics and Applied Mathematics Unit, Indian Statistical
Institute, 203 Barrackpore Trunk Road, Kolkata-700 108, India}

\author{Santanu K. Maiti}

\email{santanu.maiti@isical.ac.in}

\affiliation{Physics and Applied Mathematics Unit, Indian Statistical
Institute, 203 Barrackpore Trunk Road, Kolkata-700 108, India}

\author{Shreekantha Sil}

\affiliation{Department of Physics, Visva-Bharati, Santiniketan, West
Bengal-731 235, India}

\begin{abstract}

A new proposal is given to achieve high degree of magnetoresistance (MR) 
in a magnetic quantum device where two magnetic layers are separated by 
a non-magnetic (NM) quasiperiodic layer that acts as a spacer. The NM spacer 
is chosen in the form of well-known Aubry-Andr\'{e} or Harper (AAH) model 
which essentially gives the non-trivial features in MR due to its gaped
spectrum and yields the opportunities of controlling MR selectively by 
tuning the AAH phase externally. We also explore the role of dephasing 
on magnetotransport to make the model more realistic. Finally, we illustrate
the experimental possibilities of our proposed quantum system.
	
\end{abstract}

\maketitle

The study of magnetization dynamics where charge current is controlled by
means of magnetization configuration continues to draw venerable attention
over last few decades, and interest rapidly jumped up following the discovery
of the novel giant magnetoresistance (GMR) effect. In the late $80$'s two 
famous scientists, Albert Fert and Peter Gr\"{u}nberg, showed that a large 
change in
resistance takes place through a multilayered structure upon the
application of a magnetic field~\cite{GMR1,GMR2,GMR3,GMR4}.
It has widespread applications in designing hard disk drives, memory chips,
magnetic field dependent sensors and to name a few. Thanks to the thin film
deposition technique since without its much progress it would never have been
possible to fabricate multilayered thin film with almost a monolayer
precision for investigating the GMR effect. Though nowadays some other
structures are also available like granular 
material~\cite{GranularMolecule,grn1},
spin valve~\cite{SpinValve,spv1}, pseudo spin valve~\cite{psu}, etc., 
that can exhibit giant magnetoresistive effect. In granular
thin film some magnetic moments are randomly oriented and by applying a
magnetic field they are suitably aligned. Whereas for the other two cases,
viz, spin valve and pseudo spin valve, orientation of one magnetic layer is
changed in presence of magnetic filed though the mechanism is slightly
different because of the structural policy. Now in all these cases the basic
principle is that a large change in resistance ($\sim 10$ to $70\%$)
takes place upon the application of saturation magnetic field. Analogous to
GMR, there exists another phenomenon, known as CMR~\cite{CMR1,CMR2,CMR3},
where a huge change of
resistance takes place though its application is highly limited mainly because
of the fact that it requires very strong magnetic field ($\sim$ several
Tesla).

In magnetoresistive study main attention is being paid on how to get maximum
change in resistance due to the application of magnetic field. The greater 
change in resistance between parallel and anti-parallel configurations of two
consecutive layers in a multi-layered structure is the primary requirement 
for large data storage, and at the same time it allows to read the higher and
lower resistance states easily. In absence of any 
magnetic field when magnetization directions are different in a system, be it
a multi-layered structure or a granular material, maximum scattering of charge 
carriers takes place resulting a maximum resistance. Whereas, in presence of 
saturation magnetic field the system offers a minimum resistance. These features 
are well established with considerable theoretical and experimental
works~\cite{TheoExp1,TheoExp2,TheoExp3,TheoExp4,TheoExp5}. Considering all the 
propositions available so far in literature, a question may arise that {\em can 
we think of a device which on one hand will be very small in size, geometrically 
simple and easy to fabricate, and on the other hand, may exhibit a large 
magnetoresistanc (in some cases it may reach up to $100\%$) 
at multiple bias windows.} The $100\%$ change in MR will be 
obtained when finite propagation of
charge carriers takes place for one configuration of the magnetic layers,
while the charge flow gets perfectly blocked for the other configuration.
\begin{figure}[ht]
{\centering \resizebox*{8.25cm}{1.5cm}{\includegraphics{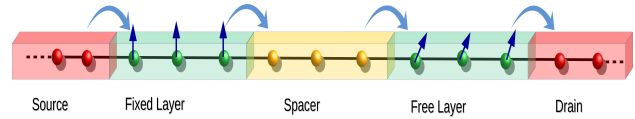}}\par}
\caption{(Color online). Schematic diagram of the nano-junction to explore
magneto-resistive effect, where a fixed magnetic layer and a free magnetic
layer is separated by a non-magnetic spacer (light yellow region). The left
and right ends of the magnetic and non-magnetic layers are connected with
semi-infinite non-magnetic perfect electrodes.}
\label{fig1}
\end{figure}
In addition, we want to tune MR externally, without applying any magnetic 
field. If this kind of device is implemented, which has not been explored 
so far, then definitely it will boost the magnetoresistive applications in 
different aspects. The present work essentially focuses on that direction.

We substantiate our proposal with the junction set-up given in 
Fig.~\ref{fig1}, where the NM layer plays the central role. One key idea 
is that we need to select the spacer in such a way that the bridging
magnetic-non magnetic-magnetic (M-NM-M) system exhibits multiple energy 
bands and they are arranged, in energy scale, differently with the parallel
(P) and anti-parallel (AP) configurations of the magnetic layers. Under
this situation $100\%$ MR can be obtained by selectively 
choosing the Fermi energy of the system, and this is one of our primary 
requisites. The other pivotal requirement is that the MR can be tuned externally.
Both these two conditions will be fulfilled with the help of an AAH
spacer~\cite{AAH1,AAH2,AAH3,AAH4,AAH5,AAH6,AAH7},
a quasicrystal, which has been a classic example of gaped systems.
Quasicrystals are found to exhibit several non-trivial topological phenomena 
that are being considered as newly developed paradigms in the discipline of 
condensed matter physics. The diverse characteristic features of AAH models
make them truly unique over the other quasicrystals, and several spectacular
phenomena have already been revealed considering both the diagonal and/or
off-diagonal versions through a reasonably large amount of recent theoretical 
and experimental works~\cite{AAH1,AAH2,AAH3,AAH4,AAH5,AAH6,AAH7}. The AAH phases 
associated with the diagonal and off-diagonal parts, those are tuned externally 
and independently, regulate the energy band structure significantly, and thus, 
tunable physical properties are naturally expected.

To make the proposed model more realistic we include the effects of 
dephasing~\cite{AAH7,Deph1,Deph2,Deph3,Deph4}.
It is an important factor that can destroy the phase memory of charge carriers,
and thus, it can affect the transport properties. Among many sources the most 
probable one is the electron-phonon (e-ph) interaction. Now inclusion of this
effect has always been a challenging task, and although some prescriptions 
are available, most of them are based on density functional theory (DFT) 
\begin{figure}[ht]
{\centering \resizebox*{7.5cm}{4cm}{\includegraphics{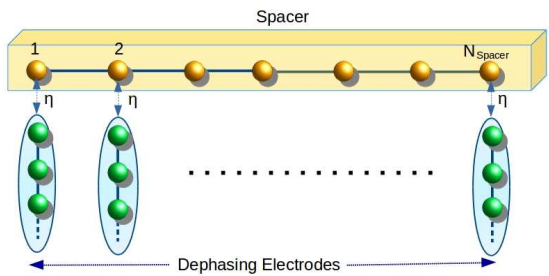}}\par}
\caption{(Color online). Arrangements of dephasing electrodes 
in the spacer region where each site of the NM spacer is directly coupled 
to the dephasing electrodes with the coupling constant $\eta$. This coupling
parameter $\eta$ is commonly known as dephasing strength. These electrodes do 
not carry any net current, but they are responsible to randomize the phases of 
the electrons. $N_{spacer}$ represents the total number of lattice sites in 
the NM spacer.}
\label{newfig1}
\end{figure}
within a non-equilibrium Green's function (NEGF) formalism which are too 
heavy to implement properly and also very time taking~\cite{dft1}. But 
B\"{u}ttiker came up with a simple and elegant idea to analyze the effect 
of dephasing~\cite{dp1,dp2,dp3}, where virtual electrodes (voltage) are 
connected at each lattice sites of the bridging system (for illustration, 
see Fig.~\ref{newfig1}) those do not carry 
any net current, but they are responsible for randomizing the phases. As
this is a classic way to include the effect of different dephasing mechanisms
we incorporate it in our present analysis. The main motivation for the 
consideration of this effect is to test whether the phenomena studied here 
still persist even in presence of dephasing or not. If they persist, then 
we will have a suitable hint that the proposed model can be tested in 
laboratory under different realistic situations.
\begin{figure*}[ht]
{\centering \resizebox*{6cm}{7.75cm}{\includegraphics{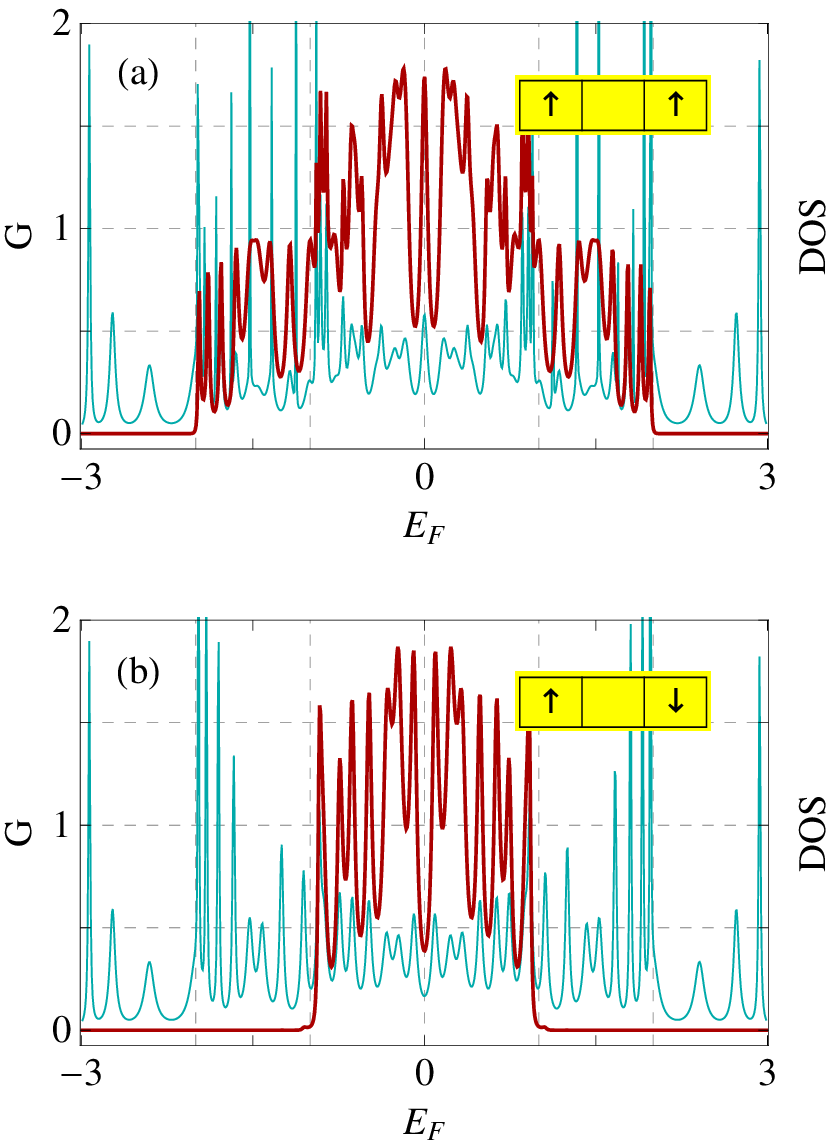}}
\resizebox*{11cm}{7.75cm}{\includegraphics{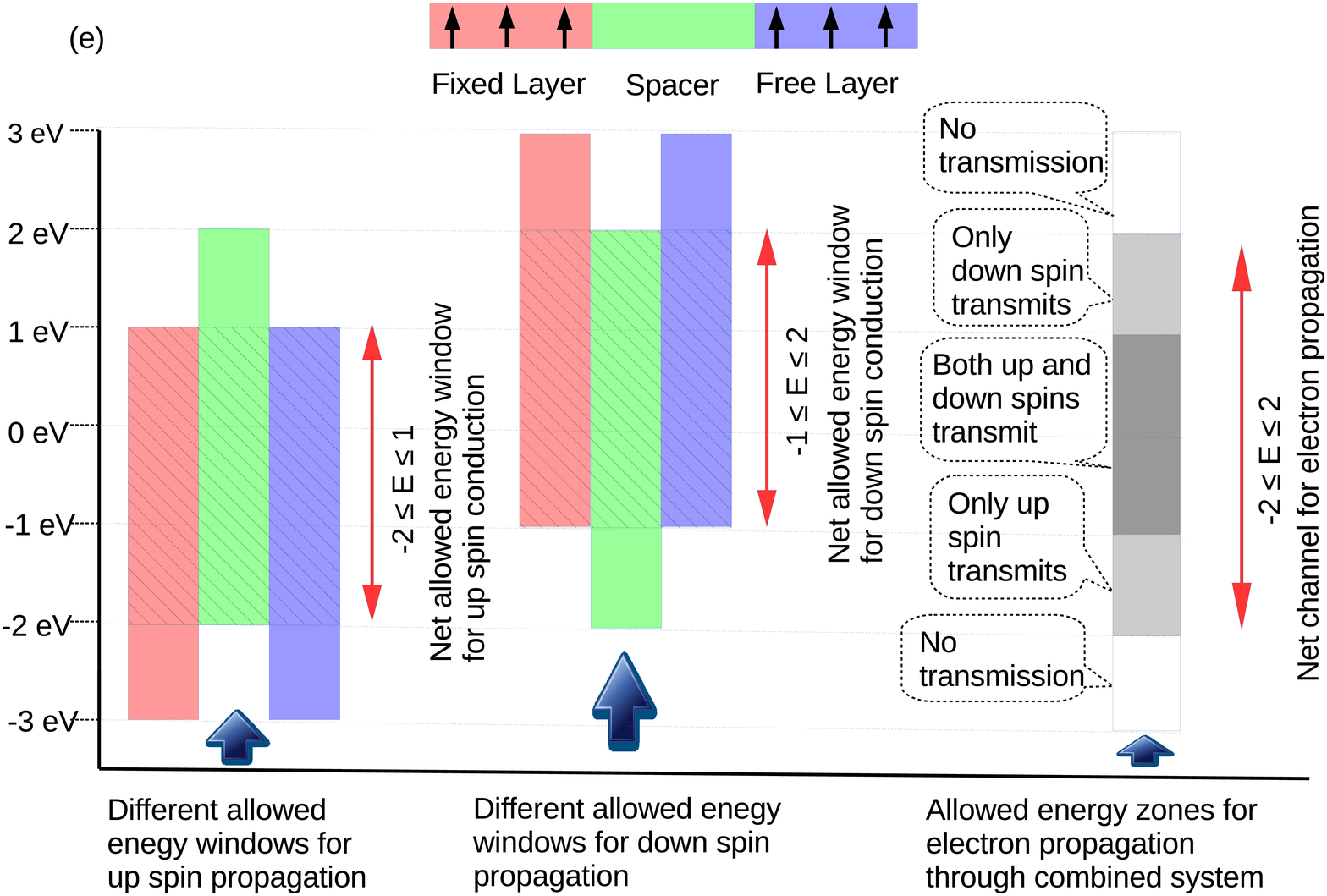}}
\resizebox*{6cm}{7.75cm}{\includegraphics{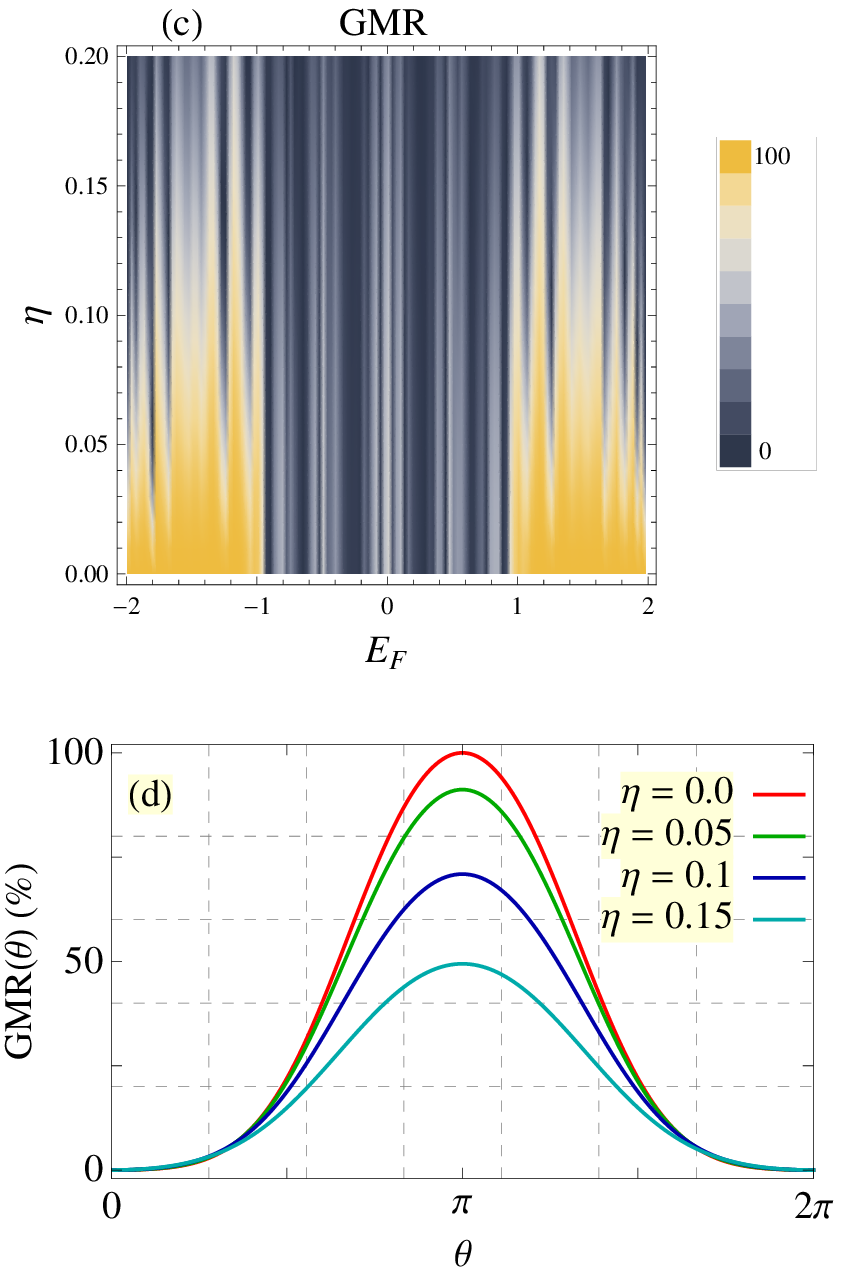}}
\resizebox*{11cm}{7.75cm}{\includegraphics{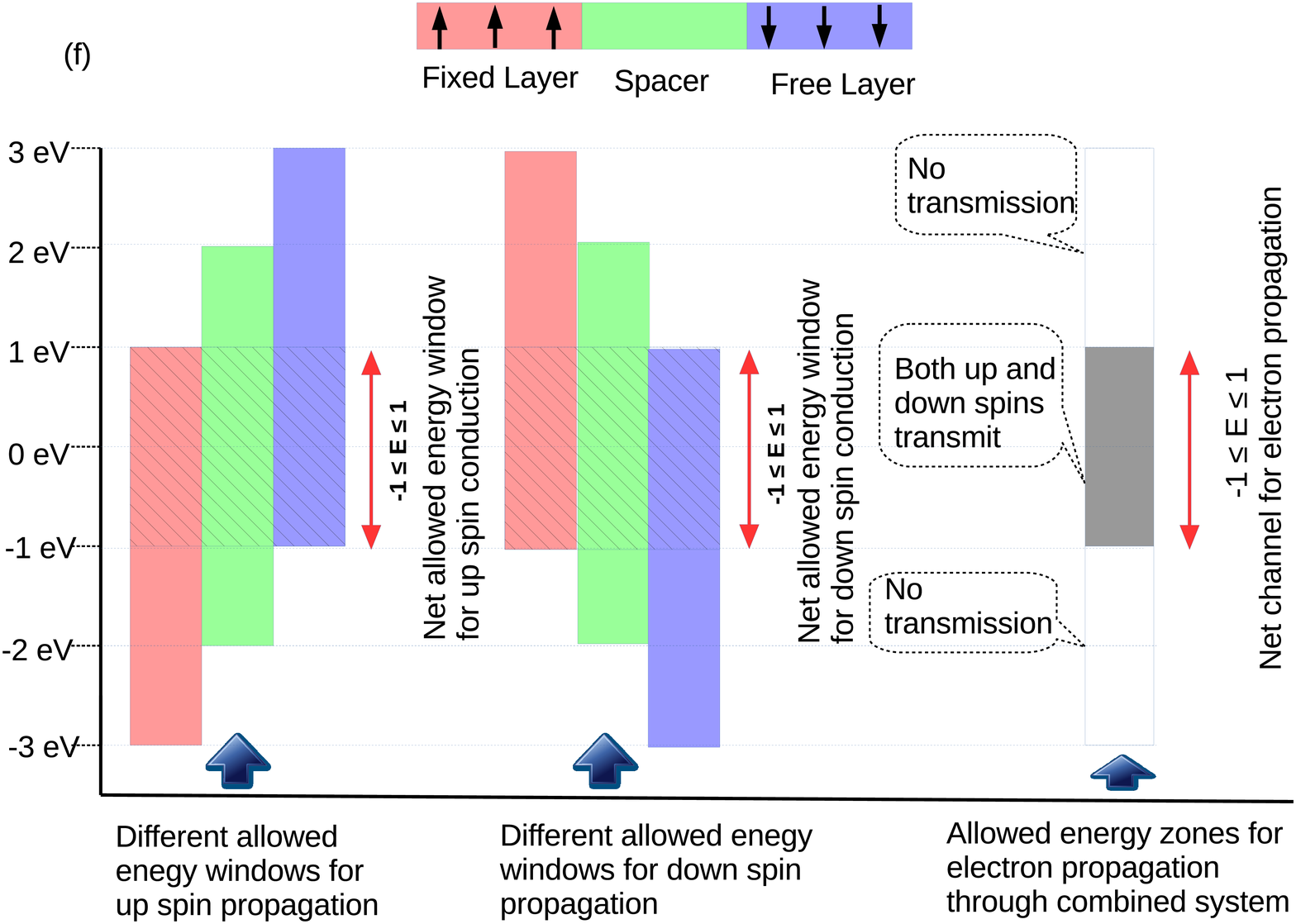}}\par}
\caption{(Color online). Illustration of GMR in presence of a perfect spacer 
($W=0$). In (a) and (b), the dependence of net conductance $G$ (red line) as 
a function of Fermi energy is shown for the parallel and anti-parallel 
configurations, respectively, and in each of these two spectra density of 
states (DOS) (sky blue color) is superimposed. In (c), a density plot is given 
to describe the simultaneous variation of GMR with dephasing 
strength $\eta$ and Fermi energy $E_F$. 
The effect of $\theta$ (we set $\theta_i=\theta\,\forall\,i$) of the free 
magnetic layer (as the magnetic moments in the other layer are always fixed
and aligned along $+Z$ direction) on GMR is presented in (d) for some typical 
dephasing strengths $\eta$. The maximum change is obtained for $\theta=\pi$, 
as expected. Finally, in (e) and (f) the allowed energy windows for the 
three different layers are drawn, to have a complete idea of transmitting 
zones for the two spin configurations. Unless otherwise stated we take only
two different values of $\theta$ for the free layer to get the parallel and
anti-parallel configurations. The other common parameter values are: 
$\epsilon_0=0$, $t_0=2$, $\epsilon_{n\uparrow}=\epsilon_{n\downarrow}=0$,
$h=1$, $N_{fixed}=N_{free}=10$ and $N_{spacer=20}$.}
\label{fig2}
\end{figure*}

We define GMR as $(G_P-G_{AP})/(G_P + G_{AP})$, where $G_P$ and $G_{AP}$
correspond to net conductances for the parallel and anti-parallel spin 
configurations, respectively. 
Usually GMR is referred as $(G_P-G_{AP})/G_{AP}$. In this 
definition a situation may arise especially for the ballistic case, which
can be understood from our forthcoming analysis, that $G_{AP}$ drops almost to
zero or in some cases it may vanish completely. Under this situation an infinite
GMR will be obtained that we may call as {\em absolute} GMR, which cannot be 
shown in the graph. To avoid this, in our work, we mention GMR as
$(G_P-G_{AP})/(G_P + G_{AP})$, and with this definition no physics will be altered.
If we get $100\%$ change in GMR, it means absolute GMR according to the other 
expression. In presence of environmental dephasing and other factors, though we
get reasonable GMR, absolute GMR cannot be obtained which can also be 
understood from our upcoming discussion.

In order to calculate GMR, we need to determine conductance and we evaluate it
from the spin dependent transmission probabilities, $T_{\sigma \sigma^{\prime}}$, 
following the Landauer conductance formula~\cite{green1}
$G_{\sigma \sigma^{\prime}}=(e^2/h) T_{\sigma \sigma^{\prime}}$, where
$\sigma (\sigma^{\prime}) = \uparrow, \downarrow$. All these components,
$T_{\sigma \sigma^{\prime}}$, are computed using the non-equilibrium 
Green's function (NEGF) formalism, which is the most suitable and standard
technique to study transport properties. In this formulation, an effective
Green's function is formed by incorporating the effects of contact electrodes
through self-energy corrections and it can be written as~\cite{green1}:
$G^r=(G^a)^{\dagger}=\left(E-H_c-H_S-H_D-\Sigma_S-\Sigma_D\right)^{-1}$,
where $\Sigma_S$ and $\Sigma_D$ are the contact self-energies, and $H_S$
and $H_D$ are the tunnel Hamiltonians due to source (S) and drain (D).
$H_c$ is the Hamiltonian of the bridging conductor which is a sum
$H_c=H_M + H_{NM}$, where $H_M$ and $H_{NM}$ are the Hamiltonians associated
with the magnetic and non-magnetic parts, respectively. We describe all
these Hamiltonians within a tight-binding (TB) framework. Using the above
Green's function we evaluate spin dependent transmission probabilities 
through the Fisher-Lee relation~\cite{green2} 
$T_{\sigma \sigma^{\prime}}^{\mbox{\tiny{SD}}}=\mbox{Tr} \left[\Gamma_{\mbox{\tiny{S}}}^{\sigma} 
G^r \Gamma_{\mbox{\tiny{D}}}^{\sigma^{\prime}} G^a \right]$, where 
$\Gamma_q^{\sigma}$'s ($q=\mbox{\tiny{S}},\mbox{\tiny{D}}$) are the
coupling matrices.

The spin dependent scattering mechanism exists only in the magnetic layers,
separated by a NM spacer, and considering this effect the TB Hamiltonian
of the magnetic layer reads as~\cite{ham1,ham2,mp1} 
$H_{\mbox{\tiny{M}}}=\sum_n {\bf c_n^{\dagger} (\epsilon_n -
h_n.\sigma) c_n} + \sum_n \left({\bf c_{n+1}^{\dagger} t c_n} + h.c. \right)$,
where ${\bf c_n^{\dagger}}$, ${\bf c_n}$ are the Fermionic operators, and
$\mathbf{\epsilon_n}$ and $\mathbf{t}$ are the ($2 \times 2$) diagonal 
matrices associated with site energy ($\epsilon_{\uparrow}$, 
$\epsilon_{\downarrow}$) and nearest-neighbor hopping (NNH) integral ($t$) 
of up and down
spin electrons. $\mathbf{h_n.\sigma}$ is the spin dependent scattering 
term where $h_n$ is the strength of magnetic moment at site $n$ and 
$\vec{\sigma}$ ($=\sigma_x, \sigma_y, \sigma_z$) is the Pauli spin vector 
with $\sigma_z$ in diagonal representation. The orientation of any magnetic 
moment is described by the usual polar angle $\theta_i$ and azimuthal angle
$\varphi_i$ in spherical polar co-ordinate system. For the NM spacer a 
similar kind of TB Hamiltonian, apart from the term $\mathbf{h_n.\sigma}$, 
is used. Now, in presence of AAH modulation, the site energy of the spacer 
becomes~\cite{AAH5}
$\epsilon_{n\uparrow}=\epsilon_{n\downarrow}=W \cos(2 \pi b\, n + \phi)$,
where $W$ is the strength of modulation and $b$ is a constant factor that
can be a commensurate or an incommensurate one. For the incommensurate AAH
model we choose $b$ as the golden mean i.e., ($1+\sqrt{5}$)/2. The other 
physical parameter $\phi$
in the site energy expression, the so-called AAH phase, plays an important 
role and it can be tuned externally with suitable set-up~\cite{AAH2,AAH5}. 
We will critically examine its effect on GMR. 

The TB Hamiltonians for the source and drain read as
$H_{\mbox{\tiny{S}}}=H_{\mbox{\tiny{D}}}=\sum_n {\bf a_n^{\dagger} 
\epsilon_0 a_n} + \sum_n \left({\bf a_{n+1}^{\dagger} t_0 a_n} + 
h.c. \right)$, where different terms correspond to the usual meanings.
These electrodes are coupled to the bridging system via the coupling
parameters $\tau_{\mbox{\tiny{S}}}$ and $\tau_{\mbox{\tiny{D}}}$, 
respectively. We assume S and D as perfect, semi-infinite, 
one-dimensional and non-magnetic.

In order to include dephasing effect following the B\"{u}ttiker prescription 
we need to couple virtual electrodes, similar to real electrodes, at each 
lattice site of the conductor (see Fig.~\ref{newfig1}). All these electrodes
are parametrized identically with S and D, and they are non-magnetic. 
The coupling strength (also referred as the dephasing strength) 
between the spacer and the dephasing electrodes is described by the parameter 
$\eta$. Now, to have the condition that these 
electrodes are not carrying any finite current, we have to adjust potentials 
($V_m$) of the virtual electrodes accordingly, such that the voltage drop 
across each of these electrodes is perfectly zero. That is in principle 
possible with the application of a finite bias across the contact electrodes 
S and D i.e., $V_S=V_0$ (say) and $V_D=0$. Under this situation, the effective 
spin dependent transmission probability is expressed as~\cite{dp3}:
$T_{\sigma \sigma^{\prime}}^{eff}=T_{\sigma \sigma^{\prime}}^{\mbox{\tiny SD}}
+ \sum_m T_{\sigma\sigma^{\prime}}^{\mbox{\tiny mD}} V_m/V_0$.

Before analyzing the results let us mention the values of the physical 
parameters those are common throughout the calculations. The on-site 
energies for the perfect lattice sites are chosen as zero, and they are 
same for both up and down spin electrons. The NNH integral, $t_0$, in S and D 
is fixed at $2$, and the other NNH integrals along with contact-to-conductor 
coupling strength i.e., $t$, $\tau_{\mbox{\tiny{S}}}$ and $\tau_{\mbox{\tiny{D}}}$, 
are set at $1$. As the dephasing strength $\eta$ is not common for all figures,
we specify it in the appropriate places during our analysis. The strength 
of magnetic moments $h$ ($h_n=h\,\forall\,n$) 
and the azimuthal angle $\varphi$ ($\varphi_n=\varphi\,\forall\,n$) 
are fixed at one and zero, respectively. The number of sites 
in the fixed and free magnetic layers are referred as $N_{fixed}$ and 
$N_{free}$, and we set them at $10$. On the other hand, for the NM spacer 
we specify the total number of atomic sites by $N_{spacer}$, and unless 
specified otherwise, we set it at $20$. All the 
energies are measured in unit of electron volt (eV).

Now we explain our results. As already stated, our central focus is 
to achieve a high degree of GMR and its suitable tuning. Before describing
the tuning mechanism, let us start to analyze how to get high GMR. The key
concept of getting high GMR is that we need to achieve higher conductance 
for one configuration of the free layer, and most importantly much lower 
conductance in the other configuration. If this lower conductance drops 
exactly to zero, then $100\%$ change in MR will be obtained. 
This can be achieved considering the layered structure as illustrated in 
Fig.~\ref{fig2}. For the parallel configuration, finite conductance is
obtained within the range $-2 \leq E_F \leq 2$ (red line of 
Fig.~\ref{fig2}(a)), whereas spin transmission gets perfectly blocked for 
both up and down spin electrons within the ranges $-2 \leq E_F \leq -1$ 
and $1 \leq E_F \leq 2$ in the anti-parallel configuration (red line of 
Fig.~\ref{fig2}(b)). Thus, setting the Fermi energy anywhere within these 
two zones, viz, $-2 \leq E_F \leq -1$ and $1 \leq E_F \leq 2$, $100\%$
GMR will be noticed. The allowed and the forbidden zones of different
spin electrons for the two different configurations of magnetic moments 
can be understood from the energy bar diagrams shown in Figs.~\ref{fig2}(e)
and (f). The electron can transmit through the junction only when a {\em common
energy channel is found}. What we see is that, for the parallel configuration
one can get finite transmission, due to up or down spin electron, in the range 
$-2 \leq E \leq 2$, among which $-1\leq E\leq 1$ is the overlap region for both 
the two spin electrons. This scenario is exactly reflected in the spectrum 
Fig.~\ref{fig2}(a). When the magnetic moments of the free layer get flipped 
to make an anti-parallel
configuration, the situation becomes more interesting. From the energy 
bar diagram Fig.~\ref{fig2}(f) we can see that only within the range
$-1\leq E\leq 1$ both the up and down spin electrons can propagate, while all 
other zones are blocked. This is the key advantage of a layered structure.
More and more selective transmitting zones can be generated by combining
more number of magnetic and NM spacers, which we check through our detailed
calculations, and thus more controlled transmission will be obtained. 
Comparing the spectra given in Figs.~\ref{fig2}(a) and (b) it is now clear 
that $100\%$ change in resistance can be possible by 
selectively choosing the Fermi energy.

The effect of dephasing is quite interesting. From the simultaneous 
variation of GMR with $\eta$ and $E_F$ (Fig.~\ref{fig2}(c)), we can see 
that for a reasonable dephasing strength a high degree of GMR is obtained. 
With increasing $\eta$ it gradually decreases, and eventually drops to zero 
for large enough strength, as expected.

Now, to examine the role of $\theta$ on GMR, in Fig.~\ref{fig2}(d) we plot
GMR as a function of $\theta$ (which we call as GMR($\theta$)) at some 
typical values of dephasing strength. For each $\eta$, the change in 
resistance increases with $\theta$, and it reaches to a maximum when all
the magnetic moments of the free layer are completely aligned in the opposite
direction with respect to the fixed layer yielding maximum scattering. 
This phenomenon leads to an important message along with the magnetoresistance
that selective spin dependent electron transport can be achieved by
changing the orientation of magnetic moments uniformly in one segment of the
junction.

The results analyzed so far are worked out considering a 
perfect spacer ($W=0$). Now, keeping in mind the {\em possible engineering 
of GMR}, we replace the perfect
\begin{figure}[ht]
{\centering \resizebox*{7.75cm}{7cm}{\includegraphics{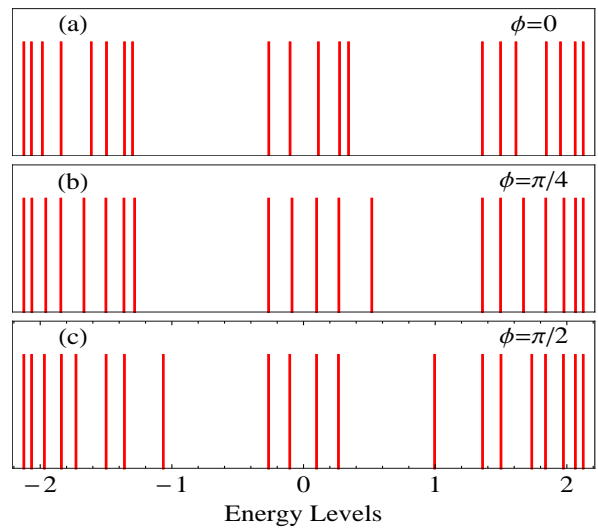}}\par}
\caption{(Color online). Energy band diagram of an 
incommensurate AAH spacer considering $N_{spacer}=20$ and $W=1$, where 
(a), (b) and (c) correspond to $\phi=0$, $\pi/4$, and $\pi/2$, respectively. 
We draw vertical lines of equal heights at each eigenvalues of the AAH spacer 
to have the full energy spectrum. The gaped nature along with the tuning of 
energy levels with the AAH phase $\phi$ are clearly visible.}
\label{newfig2}
\end{figure}
spacer by the AAH one. The speciality of the AAH spacer is that it exhibits 
gaped spectrum, as clearly seen from Fig.~\ref{newfig2} where the energy 
levels are plotted for a $20$-site incommensurate AAH chain at three typical
values of $\phi$. In addition to that, the possible tuning of energy spectrum 
\begin{figure*}[ht]
{\centering
\resizebox*{8.5cm}{6.5cm}{\includegraphics{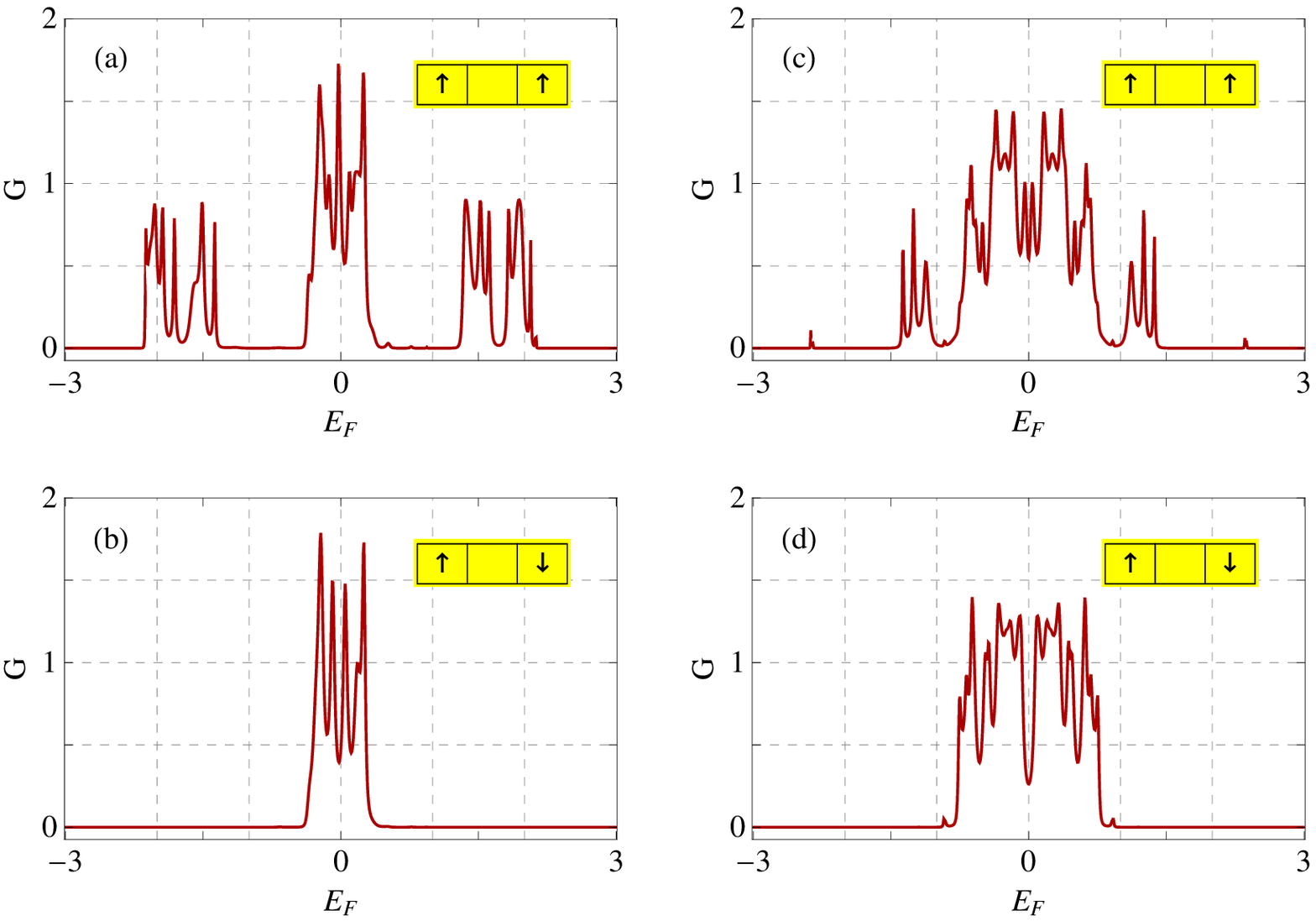}}
\resizebox*{5.5cm}{6.5cm}{\includegraphics{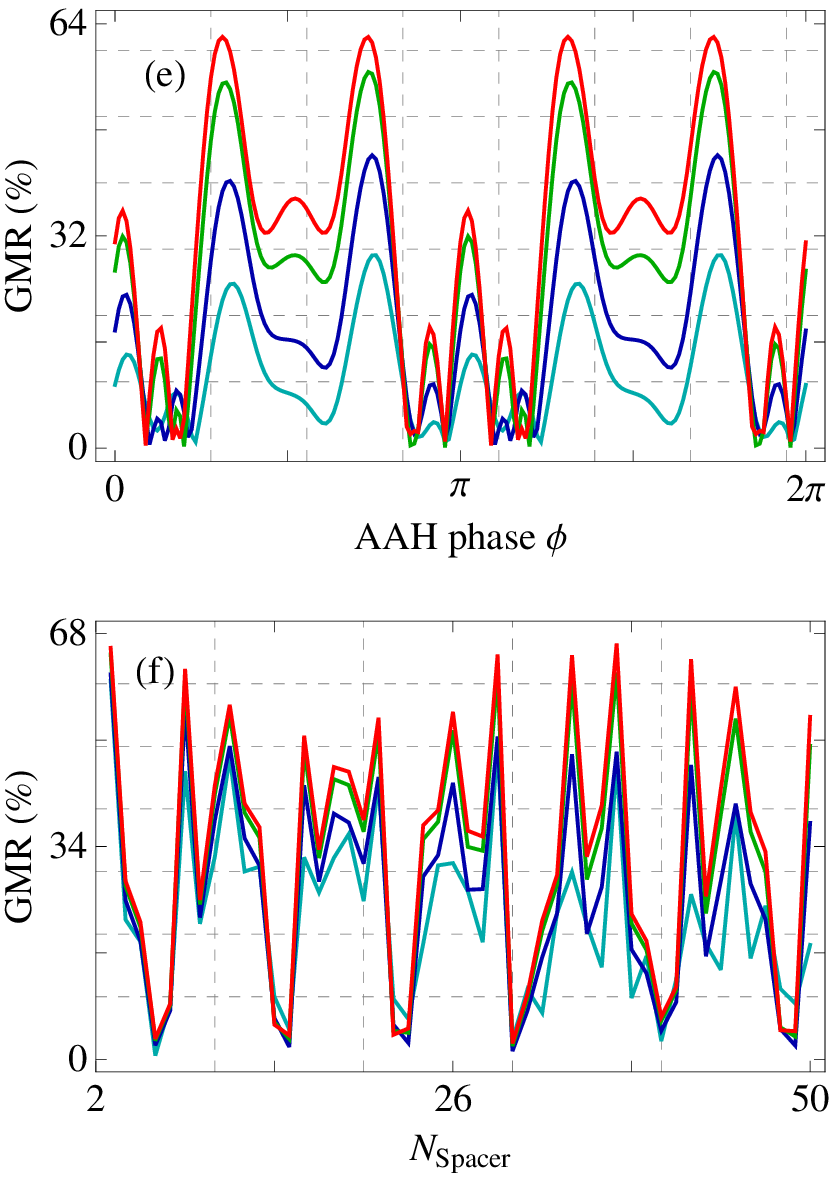}}\par}
\caption{(Color online). Implementation of GMR in presence of an AAH spacer.
In (a)-(d), net conductance $G$ is shown as a function of Fermi energy $E_F$
for both the two orientations of magnetic moments, where the results 
in (a) and (b) are given for the incommensurate ($b$=golden mean) AAH spacer,
and in (c) and (d) the results are presented for the commensurate ($b=1/10$) 
one. To compute these results, we set the AAH modulation strength $W=1$ and 
the phase $\phi=0$. Comparing the conductance spectra, the appearance of high 
degree of GMR at selective Fermi energies can easily be understood. The role 
of phase $\phi$ on GMR is illustrated in (e) considering an incommensurate AAH 
spacer with $W=1.5$, and finally, in (f) the effect of the spacer size is 
established where we fix $W=1$ and $\phi=0$. The different colors in (e) and 
(f) correspond to the results for the identical dephasing strengths as taken in 
Fig.~\ref{fig2}(d). For (a)-(e) we choose $N_{spacer}=20$. The other common 
parameter values are: $\epsilon_0=0$, $t_0=2$, $h=1$, $N_{fixed}=N_{free}=10$ 
and $\theta=0$.}
\label{fig3}
\end{figure*}
is also clearly reflected, which is another key advantage of AAH lattices
compared to other conventional uncorrelated and correlated disordered ones.
Because of the gaped spectrum exhibited by the AAH spacer, several energy 
zones are expected where finite transmission is available from one 
configuration (parallel), while almost zero contribution is obtained for 
the other configuration (anti-parallel). This is exactly reflected from 
the spectra given in Figs.~\ref{fig3}(a) and (b), where the conductance 
is shown for an incommensurate AAH spacer. The zero contribution in 
conductance for the anti-parallel arrangement is due to the non-availability 
of common spin channel, as clearly discussed earlier in the case of perfect 
spacer. Multiple energy windows are available where conductance drops exactly 
zero in the anti-parallel configuration resulting a $100\%$ change in MR.
Several such energy windows cannot be observed in the case of a perfect 
spacer. For the sake of completeness, in Figs.~\ref{fig3}(c) and (d) we 
present the results of a commensurate AAH spacer, and comparing the 
spectra given in Fig.~\ref{fig3} we can clearly emphasize that the 
incommensurate AAH spacer is superior than the commensurate one. Here we
would like to note that, for commensurate $b$ the system becomes a perfect
one which exhibits always extended energy eigenstates. Whereas, for the
incommensurate $b$ the system becomes a correlated disordered one which
thus exhibits non-trivial energy spectrum rather than a commensurate one.
Because of this fact, interesting behavior in transport phenomena is naturally
expected for incommensurate AAH system.

From the spectra shown in Fig.~\ref{newfig2} it is now clear that AAH phase
$\phi$ has a critical role in energy band engineering, which thus definitely
be reflected in GMR effect. Now we analyze it clearly. In the energy regions 
where finite conductance is obtained for one configuration and the 
conductance becomes zero for the other configuration, always $100\%$ GMR 
will be obtained, and hence there will be no meaning to examine the effect 
of AAH phase in those energy zones. Therefore, we select $E_F$ in such a 
way where conductance is finite for both the two configurations of the 
magnetic layers. Here we set $E_F=0$. From the results given in 
Fig.~\ref{fig3}(e) we find that a reasonably large change in GMR is 
obtained by tuning the phase factor $\phi$ for each dephasing strength 
$\eta$ which leads to an important conclusion that one can selectively 
choose the phase $\phi$ to achieve higher GMR, and most importantly, it 
can be performed externally~\cite{AAH2,AAH5}. In this 
context it is relevant to note that few other proposals have also been 
made in different set-ups for possible tuning of MR 
{\em externally}~\cite{kmr1,kmr2,kmradd1}. For instance, considering a graphene 
heterostructure Bala Kumar {\em et al.}~\cite{kmr1} have shown that a large 
MR can be achieved upon the application of magnetic field, employing the 
specific properties of wave functions in the field and zero-field cases. 
On the other hand, in another work by Bala Kumar and co-workers~\cite{kmr2} 
it has been established that band engineering can be possible in graphene 
nanoribbon by applying external magnetic field which leads to a large 
change in MR. With these proposals we get a clear confidence for 
implementing a new prescription of externally controlled mechanism of MR. 
Our work, thus definitely a new addition along this line.

Finally, to test how the results are sensitive to the size of the AAH spacer,
in Fig.~\ref{fig3}(f) we plot GMR by varying the length of the spacer 
considering $\phi=0$ and $E_F=0$. It exhibits pronounced oscillations
providing almost constant amplitude with $N_{spacer}$, which gives us a
hint of choosing the dimension of the spacer for better performance.

From the results studied here we see that both for the 
ordered and AAH spacers, though GMR gets reduced with dephasing strength 
$\eta$, high degree of GMR can still be observed for a reasonably large
$\eta$. This is one way (means the inclusion of dephasing) to include 
the environmental effects/disturbances, as put forward by B\"{u}ttiker
and many other groups. At the same time another few factors are also 
there that may affect the GMR. For instance, bulk disorder and/or edge 
vacancies, depending on the specific geometry of the conducting junction.
It is true that disorder modifies the transport 
properties~\cite{kmr3,yao,gangu}, but as GMR
\begin{figure}[ht]
{\centering \resizebox*{8cm}{4.75cm}{\includegraphics{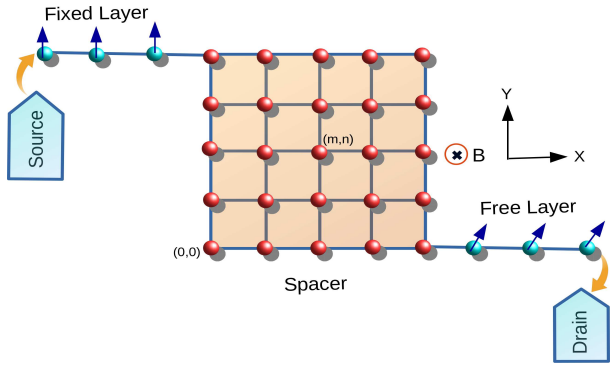}}\par}
\caption{(Color online). Proposed set-up to realize the model experimentally.}
\label{expt}
\end{figure}
is the ratio between two conductances, significant change in GMR will not
be noticed even for moderate disorder strength, which can also be confirmed
from our results considering the AAH spacer (AAH model is called as the
correlated disordered one). For strong enough disorder, when the states 
are almost localized, naturally we cannot expect any such phenomena.

Now, considering the unique and diverse characteristic 
features of AAH lattices, one proposition may come to our mind that instead 
of using an AAH spacer between two magnetic layers can we think about a 
GMR set-up where the effect of AAH potential is directly implemented into
the magnetic layers, removing the spacer region. Of course the opportunity of
energy band engineering will be still there by changing the AAH phase, but
in the absence of NM spacer, two magnetic layers will then interact with each 
other because of the magnetic exchange interaction among them, which 
essentially affects the magnetic layers. To avoid this magnetic interaction, 
the inclusion of a NM layer is highly recommended, as used in other GMR 
studies.

For experimental realization of our proposed model, we can think about a 
set-up given in Fig.~\ref{expt} where two magnetic layers are separated by 
a 2D lattice subjected to a transverse magnetic field $B$, the so-called
quantum Hall system. It is well-known that a 2D Hall system maps 
exactly to an effective 1D chain where the site energy gets modulated 
with the factor $B$. Thus, selectively tuning the magnetic field one
can design a spacer in the form of AAH chain, and in principle, can 
examine the results studied here.

In conclusion, we have established a new proposal to achieve better 
performance in magnetorestive effect exploiting the unique features
of correlated disordered lattice, that has not been reported so far
in literature to the best of our knowledge. The persistence of the 
results even in presence of large dephasing strength gives us a confidence 
that the proposal can be substantiated experimentally with suitable set-up. 
What comes out from the entire analysis is that the essential
mechanism of magnetoresistance is hidden within the non-trivial 
characteristics of different spacers, and here we have shown one example
along this direction considering an AAH system. We also get a strong 
confidence about our claim following one recent work done by Wang {\em
et al.}~\cite{wang}. Considering a bottom-pinned perpendicular 
anisotropy-based magnetic tunnel junction (p-MTJ, stacked with Tungsten (W)
layers and MgO/CoFeB interfaces, they have shown that a large 
magnetoresistance $\sim 249\%$ can be achieved, circumventing Tantalum 
(Ta) as the spacer as was used previously in other p-MTJ films. Thus, 
undoubtedly the spacer has the most significant role in magnetoresistive
study. Although several propositions have been put forward, still more 
investigations are required for better performance.

SKM would like to thank the financial support of DST-SERB, India under 
Grant No. EMR/2017/000504.

\end{document}